\documentstyle [12pt,a4fedele] {article}
\begin{document}
\hfuzz=50pt
%
%
\newcommand{\be}{\begin{equation}}
\newcommand{\ee}{\end{equation}}
\newcommand{\bea}{\begin{eqnarray}}
\newcommand{\eea}{\end{eqnarray}}
\begin{titlepage}
\begin{center}
 {\Large \bf Constant Curvature and Non-Perturbative $W_3$ Gravity  }\\
 \vspace{2cm}
 {\large \bf Noureddine Mohammedi}\footnote{nour@itsictp.bitnet}\\
\vspace{.5cm}
\large International Centre for Theoretical Physics \\
34100 Trieste, Italy.\\

\baselineskip 18pt

\vspace{.2in}

September, 1991\\
\vspace{1cm}
Abstract
\end{center}
We show that the new classical action for two dimensional gravity
(the Jackiw-Teitelboim model) possesses a $W_3$ algebra. We quantise
the resulting $W_3$ gravity in the presence of matter fields with arbitrary
central charges and obtain the critical exponents. The auxiliary field of the
model, expressing the constancy of the scalar curvature, can be interpreted
as one of the physical degrees of freedom of the $W_3$ gravity. Our expressions
are corrections to some previously published results for this model where the
$W_3$ symmetry was not accounted for.\\

\end{titlepage}
\baselineskip 20pt

\section{Introduction}

\setcounter{footnote}{0}
\par
Remarkable  are the efforts put into the understanding of gravity [1-4] and
matrix models [5-8], yet  minute are the advances made in the construction of
realistic strings outside of critical dimensions. Any attempt to couple matter,
with central charge $c$ in the domain $1<c<25$, to two dimensional gravity
is faced with the issue that some critical exponents start aquiring complex
values. It is highly probable that this difficulty arises from our lack of
a generally covariant action for pure gravity in two dimensions.
\par
Indeed, it was shown in ref.[9] that if one assumes that the classical gravity
action is given by
\be
S_{JT}={b\over \pi}\int d^2x\sqrt g N(R+\Lambda)
\ee
then the quantised theory does allow for matter couplings with arbitrary
central charges. This action is an alternative to the non-local gravitational
functional induced by the coupling of matter fields to gravity [1]. In the
above
$b$ is a constant that can be absorbed into $N$, and $\Lambda$ is the
cosmological constant. The field $N(x)$ is an auxiliary field playing
a crucial role in our analyses and classically imposes the constancy of the
curv
   ature $R$ through its equation of motion
\be
R+\Lambda =0\,\,\,.
\ee
In the literature $S_{JT}$ is known as the Jackiw-Teitelboim action [10] and
some interesting aspects of it have been considered in [11-15]. Recently the
perturbative behaviour of this model was also analysed in [16,17].
\par
In this note we show that the Jackiw--Teitelboim model of gravity possesses
a $W_3$ algebra [18]. We succeeded in constructing, in addition to the
energy-momentum tensor, a spin-three generator as a function of the field
$N(x)$ and the Liouville mode. Since for $W_3$ gravity
\footnote{The $W_N$ gravities can be thought of as arising upon gauging
$W_N$ algebras [19].}
there are two
physical degrees of freedom, the auxiliary field $N(x)$ can be interpreted
as one of these degrees of freedom (the other one being the Liouville mode).
This $W_3$ gravity can be quantised in the presence of matter fields with
arbitrary central charge. Our results are corrections to those expressions
obtained in [9] where the $W_3$ algebra was not taken into account.
\par
We start by showing that in the conformal gauge the energy-momentum tensor
corresponding to $S_{JT}$ has a central charge equal to 2 and no $W_3$
generator exists. In section 3, the contribution arising from the measure
is taken into consideration and the explicit form of the spin-three generator
is given. We also discuss the manner in which matter fields couple to $W_3$
grav
   ity.
Finally, the critical exponents of the model are computed. The final section
is devoted to conclusions and outlook.

\section{The Classical Model}
\par
In the background geometry specified by $\hat g_{\alpha\beta}$, where
\be
g_{\alpha\beta}=\hat g_{\alpha\beta} e^{\gamma\sigma}\,\,\,,\,\,\,
\gamma=1\,\,or\,\,2\,\,\,,
\ee
the Jackiw-Tetelboim action takes the form
\be
S_{JT}={b\over \pi}\int d^2 x\sqrt{\hat g}[\gamma\hat g^{\mu\nu}\partial_{\mu}
\sigma\partial_{\nu} N + N(\hat R+\Lambda e^{\gamma\sigma})]\,\,\,,
\ee
where we have made use of the results
\bea
R&=&e^{-\gamma\sigma}(\hat R-\gamma\nabla^2_{\hat g}\sigma)\nonumber\\
\nabla^2_{\hat g}&=&{1\over\sqrt{\hat g}}\partial_{\alpha}(\sqrt{\hat g}
\hat g^{\alpha\beta}\partial_{\beta})\,\,\, .
\eea
The energy momentum tensor corresponding to $S_{JT}$, with $\Lambda=0$,
is given by
\bea
T_{\alpha\beta}&\equiv &-{4\pi\over\sqrt{\hat g}}{\delta L\over\delta\hat
g^{\alpha
\beta}}=-4b[2\gamma\partial_{(\alpha}\sigma\partial_{\beta )}N -
\hat\nabla_{\alpha}
\hat\nabla_{\beta}N\nonumber\\
& +& \hat g_{\alpha\beta}(-{\gamma\over 2}\hat g^{\mu\nu}\partial_{\mu}
\sigma\partial_{\nu}N+\hat \nabla ^\mu\hat \nabla _\mu N)]\,\,\,
\eea
and its {\it z-z} component is written as
\be
T_{zz}\equiv T(z)=-4b(2\gamma\partial_z\sigma\partial_z N-\partial^2_zN)\,\,\,.
\ee
The kinetic term in $S_{JT}$ produces the only propagator of this theory
\be
<\sigma (z) N(w)>=-{1\over 8\gamma b} \ln(z-w)\,\,\,.
\ee
Using this propagator one finds that $T(z)$ has a central charge of  2.
\par
Since for $W_3$ gravity there are precisely two physical degrees of freedom
in the form of two scalars with possible background charges, one might suspect
t
   hat there exists, in addition to the energy-momentum tensor, a $W_3$
generato
   r
for the Jackiw-Teitelboim model. To investigate this issue it is convenient
to define the  pair of scalar fields
\bea
\lambda &=&{8b\gamma\over x}\sigma + xN\nonumber\\
\rho &=&-{8b\gamma\over x}\sigma + xN\,\,\,\,,
\eea
where $x$ is any parameter different from zero. In terms of these new fields
the energy-momentum tensor is written as
\be
T(z)=-{1\over 4}(\partial_z\lambda\partial_z\lambda-\partial_z\rho
\partial_z\rho)+{2b\over x}(\partial_z^2\lambda +\partial_z^2\rho)
\ee
and we have the following operator products for the fields $\lambda$ and $\rho$
\bea
\lambda (z)\lambda (w)&=&-2\ln\,\,(z-w)\nonumber\\
\rho (z)\rho (w)&=&2\ln\,\,(z-w)\nonumber\\
\lambda (z)\rho (w)&=&0\,\,\,\,.
\eea
\par
With the form of the background charges for the fields $\lambda$ and $\rho$
as given in (2.8), we were not able to find any spin-three generator that
forms with $T(z)$ a closed $W_3$ algebra. This can be also seen from the
results of refs.[20,21].
The main result of this note is to show that such a $W_3$ algebra
{\it does} exist
when the contribution coming from the measure is taken into account.

\section{The $W_3$ Algebra}
\par
We assume, following [3,4], that the product of the measures of the fields
scale as\footnote {A similar supposition was made for $W_N$ gravity in [22].}
\bea
{\cal M}_g&\equiv& {\cal D}_gN\,{\cal D}_g\sigma\,{\cal D}_gX\,{\cal D}_gbc
\nonumber\\
&=&{\cal M}_{\hat g}\,\,\exp[-S(\sigma,\hat g)]
\eea
under the definition (2.1). Here ${\cal D}_gX$ stands for the integration over
matter fields while ${\cal D}_gbc$ corresponds to that of the $(b,c)$ ghost
system. The action $S(\sigma,\hat g)$ is of the same   form as the Liouville
action
but with arbitrary coeffecients [3,4]. This is renormalised to take the form
\be
S(\sigma,\hat g)={1\over \pi}\int d^2x\sqrt {\hat g}(\hat g^{\mu\nu}
\partial_\mu\sigma\partial_\nu\sigma+Q\sigma\hat R +\mu e^{
\gamma\alpha\sigma})\,\,\,,
\ee
where $\mu$ is the cosmological constant and the parameters $Q$ and $\alpha$
are determined by requiring the vanishing of the total conformal anomaly and
that $e^{\gamma\alpha\sigma}$ is a (1,1) conformal operator.
\par
Now the action $S=S_{JT}+S(\sigma,\hat g)$, with $\Lambda =\mu =0$,  has a
total
    energy-momentum tensor given by
\be
T(z)=-{1\over 4}(\partial_z\lambda\partial_z\lambda-\partial_z\rho
\partial_z\rho)+ia_1\partial_z^2\lambda +ia_2\partial_z^2\rho\,\,\,,
\ee
where
\bea
a_1&=&-i[{y\over 4b\gamma}+{1\over 8yb\gamma^2}(16b^2\gamma^2 -y^2)]
\nonumber\\
a_2&=&-i[-{y\over 4b\gamma}+{1\over 8yb\gamma^2}(16b^2\gamma^2 +y^2)]\,\,\,.
\eea
The new fields $\lambda$ and $\rho$ are now given by
\bea
\lambda &=&{1\over 2yb\gamma }(16b^2\gamma^2 +y^2)+yN\nonumber\\
\rho &=&-{1\over 2yb\gamma}(16b^2\gamma^2 -y^2)+yN
\eea
and their operator products are still given (2.9).
Again $y$ is a non-zero parameter.
\par
The spin-three operator, $W_{zzz}\equiv W(z)$, that completes the $W_3$ algebra
is given by
\be
W(z)=-{i\over 12} d_{ijk}:\partial_z\phi^i\partial_z\phi^j\partial_z\phi^k:
-e_{ij}:\partial_z\phi^i\partial_z^2\phi^j:+if_j\partial_z^3\phi^j\,\,\,,
\ee
where the indices $i,j$ and $k$ run from 1 to 2 and $\phi^1=\lambda$ and
$\phi^2=\rho$. For the closure of the $W_3$ algebra the quantities $d_{ijk}$,
$e_{ij}$ and $f_i$ have to satisfy certain consistency equations [20,21]. The
coefficients $d_{ijk}$ are totally symmetric while $e_{ij}$ has no
particular symmetry.  These are listed below :
\bea
d_{111}&=&d_{122}=\sqrt 2 {a_2\over\sqrt {a_1}}(a_1+3a_2)(3a_1^2+a_2^2)
/[(a_1^2+3a_2^2)\sqrt {15a_1-15a_2-4}\sqrt {(a_1-a_2)^3}] \nonumber\\
d_{112}&=&d_{222}=\sqrt 2 \sqrt {a_1}(a_1+3a_2)
/[\sqrt {15a_1-15a_2-4}\sqrt {(a_1-a_2)^3}] \nonumber\\
e_{11}&=&e_{22}=\sqrt 2 \sqrt {a_1} a_2(a_1+3a_2)(a_1^2-a_2^2)
/[(a_1^2+3a_2^2)\sqrt {15a_1-15a_2-4}\sqrt {(a_1-a_2)^3}] \nonumber\\
e_{12}&=&{\sqrt 2\over 4\sqrt {a_1}}(a_1+3a_2)(3a_1^2+a_2^2)(a_1^2-a_2^2)
/[(a_1^2+3a_2^2)\sqrt {15a_1-15a_2-4}\sqrt {(a_1-a_2)^3}] \nonumber\\
e_{21}&=&{\sqrt 2\over 4\sqrt {a_1}}(a_1+3a_2)(a_1^2-a_2^2)
/[\sqrt {15a_1-15a_2-4}\sqrt {(a_1-a_2)^3}] \nonumber\\
f_1&=&{1\over\sqrt 2}{a_2\over\sqrt{a_1}}(a_1+3a_2)(a_1^2-a_2^2)^2
/[(a_1^2+3a_2)\sqrt {15a_1-15a_2-4}\sqrt {(a_1-a_2)^3}] \nonumber\\
f_2&=&{\sqrt {a_1}\over \sqrt 2}(a_1+3a_2)(a_1^2-a_2^2)^2
/[(a_1^2+3a_2^2)\sqrt {15a_1-15a_2-4}\sqrt {(a_1-a_2)^3}]
\eea
\par
In the operator product expansion of $W(z)W(w)$, the spin-four operator
$\Lambda (z)$ is normal ordered with respect to the modes of $T(z)$ regardless
of the modes of the fields from which $T(z)$ is made [20]. However, if
$\Lambda (z)$
is normal ordered with respect to the modes of  $\lambda$ and $\rho$ then
there exist no spin-three  operator that forms a closed $W_3$ algebra [21]. It
is important to notice that the above solutions in (3.7) cannot be obtained
from
the general construction of ref.[20]. This is because there one has to impose
th
   e constraint, $(a_1^2+3a_2^2)=0$,
among the background charges. It is clear that this condition cannot be
compatib
   le
with our solution as it renders  undefined the coefficients in (3.7).
This is also in favour of the claim made in [23] that  more general solutions
ca
   n be found for the consistency equations of ref.[20].
\par
The central charge of the $W_3$ gravity is independent of the value of $y$ and
is given by
\be
c_w=2 +24({2\over \gamma}Q -{1\over\gamma^2})\,\,\,.
\ee
The ghosts mentioned in the measure (3.1) correspond to the $T$ and $W$
generato
   rs.
These are the ghost-antighost pairs $(c_2,b_2)$ and $(c_3,b_3)$ with scaling
dimensions $(-1,2)$ and $(-2,3)$ respectively. Their contribution to the
central charge is given by the general expression for $W_N$ algebras ($N=3$ in
o
   ur case),
\bea
c_{gh}(N)&=&-(N-1)(4N^2+4N+2)\nonumber\\
c_{gh}(3)&=&-100\,\,\,.
\eea
\par
When our system is coupled to some conformal field theory (matter) with
central charge $c$, the requirement that the total conformal anomaly
vanishes leads to
\be
Q={\gamma\over 48}(98 +{24\over\gamma^2}-c)\,\,\,.
\ee
This is a correction to the result of ref.[9] in which the $W_3$ symmetry
was not taken into account. The parameter $\alpha$ is determined by computing
the conformal weight of the operator $e^{\gamma\alpha\sigma}$. This is given by
\be
T(z)e^{\gamma\alpha\sigma}(w)={\alpha\over (z-w)^2}
+{1\over (z-w)}\partial_we^{\gamma\alpha\sigma}(w)\,\,\,.
\ee
Therefore $e^{\gamma\alpha\sigma}$ is a (1,1) conformal operator iff
\be
\alpha=1
\ee
\par
The question to be answered now is how does matter couple to our system ? For
simplicity we represent the matter with $d$ free scalar fields, $X^i(z)$,
without background charges, i.e. $c=d$. The total energy-momentum tensor can
then be written  as
\be
T(z)=-{1\over 4}g_{ij}:\partial_z\phi^i\partial_z\phi^j:
+ia_1\partial_z^2\lambda
+ia_2\partial_z^2\rho\,\,\,,
\ee
where $\phi^1=\lambda$, $\phi^2=\rho$ and $\phi^i=X^i$ for $i=3,\dots,d+2$.
The metric $g_{ij}$ is a $(d+2)\times (d+2)$ diagonal matrix,
$g_{ij}=diag\,\,(1,-1,1,\dots,1)$. The total spin-three generator is still
given
by the expression (3.6), where the indices $i,j$ and $k$  run from 1 to $d+2$
and the non-vanishing coefficients are given by [20]
\bea
d_{111}&=&d_{122}=\kappa\nonumber\\
d_{1ij}&=&-\kappa g_{ij}\,\,\,\,\,\,for\,\,\,\,\,\,i,j=3,\dots ,d+2
\nonumber\\
e_{11}&=&e_{22}={1\over 2}a_1\kappa ,\,\,,\,\,\,e_{12}=-a_2\kappa\nonumber\\
e_{ij}&=&-{1\over 2}a_1\kappa g_{ij}\,\,\,\,\,\,for\,\,\,\,\,\,i,j=3,\dots ,d+2
\nonumber\\
f_1&=&{1\over 3}a_1^2\kappa \,\,\,,\,\,\,f_2=-a_1a_2\kappa\,\,\,.
\eea
Here $\kappa^2=b^2\equiv 16/( 22+5\widetilde c)$ and
$ \widetilde c=d+2+24({2\over\gamma}Q-{1\over\gamma^2})$. Finally $a_1$ and
$a_2$ are given by (3.4) but with $y$ replaced by
\be
y={\pm b\gamma\over\sqrt 2(2\gamma Q-1)}[
\sqrt {48\gamma Q-d\gamma^2-24}\pm\sqrt {-16\gamma Q-d\gamma^2+8}]\,\,\,.
\ee
This is a solution to the condition imposed in [20] on the background charges
\be
a_1^2+3a_2^2=-{1\over 8}d\,\,\,.
\ee
\par
To extract out the critical exponents of the theory, one has to determine the
area dependence of the partition function $Z$. From the constancy of the scalar
curvature in (1.2) and the formula for the Euler characteristic, \newline
$\int d^2x\sqrt g R=4\pi(1-h)$, we get
\be
A\Lambda =4\pi (h-1)\,\,\,,
\ee
where $h$ is the genus of the two dimensional manifold and $A =\int d^2x\sqrt
g$
is its area. Hence $Z(A)$ is obtained by computing the $\Lambda$ dependence
of $Z$. This is achieved by considering the rescaling $\sigma
\,\,\rightarrow\,\
   ,
\sigma+{\beta\over \gamma}$ for some constant $\beta$. The total action then
shifts  by
\be
S_{tot}(\Lambda)\,\,\rightarrow\,\,S_{tot}(\Lambda e^\beta )+
{\beta\over 48}(98 +{24\over\gamma^2}-c)\int d^2x\sqrt {\hat g} \hat R
\,\,\,,
\ee
where $S_{tot}=S(\sigma,\hat g)+S_{JT}+S_{matter}+S_{ghost}$ with $\mu =0$.
The matter and the ghost action are scale invariant, as is also the measure of
the path integral. It follows that
\be
Z(\Lambda)=Z(\Lambda e^\beta )\,\,\times \,\,\exp\,\,[-
{\beta\over 12}(98 +{24\over\gamma^2}-c)(1-h)]\,\,\,.
\ee
The solution to this equation is
\bea
Z(\Lambda)&\sim & \Lambda ^{{1\over 12}(98 +{24\over\gamma^2}-c)(1-h)}
\nonumber\\
Z(A)&\sim & A ^{-1-{1\over 12}(98 +{24\over\gamma^2}-c)(1-h)}\,\,\,,
\eea
where we have used the relation between $A$ and $\Lambda$ in (3.17). The
$A^{-1}
   $
factor arises from the scaling of the delta function, $\delta (
\int d^2x\sqrt   g  R -A)$, inserted in the path integral.
The string susceptibility is defined by $Z(A)\,\sim\,A^{\gamma_{str}-3}$ and
is therefore given by
\be
\gamma_{str}=2-{1\over 12}(98 +{24\over\gamma^2}-c)(1-h)\,\,\,.
\ee
This is also a correction to the susceptibility expression of ref.[9].

\section{Conclusions}
\par
In this letter we have shown that the Jackiw-Teitelboim model of two
dimensional
gravity possesses a $W_3$ algebra at the quantum level. Matter fields
with arbitrary central charge can be coupled to it and no critical dimensions
arise. Therefore this model is a strong candidate for a theory of $W_3$ gravity
and non-critical "$W_3$ strings".
This also allows for interpreting the auxiliary field of the model as one of
the physical degrees of freedom of $W_3$ gravity. Thus, it is suggestive to
postulate that $W_3$ gravity might derive from higher dimensions upon
performing some form of dimensional reduction [24]. This is because the
Jackiw-Teitelboim model can be viewed as arising from the three dimensional
Einstein-Hilbert action when the dependence on the third dimension is
suppressed and the components of the metric in this direction are
$g_{13}=g_{23}=0$ and $g_{33}=N^2(x)$.
\par
The important question that one would like to find an answer to is how
does the non-linear $W_3$ symmetry come about  in this model ?
One way of shedding some light on this issue is by representing our model
by a two dimensional topological field theory based on some gauge group.
It is then possible to see that in order to couple matter fields to
this topological theory, one needs to introduce a rank-three tensor $B_{\mu\nu
\rho}$ that transforms like the extra gauge field of $W_3$ gravity [17]. This
line of research deserves certainly further investigation.

\par
{\bf Acknowledgements:} We would like to thank the IAEA and UNESCO for
financial support.

\end{document}